\documentclass[prl,aps,twocolumn,notitlepage,superscriptaddress,showpacs,nofootinbib]{revtex4-2}

\usepackage{enumerate,appendix}
\usepackage{amsmath, amsthm, amssymb,commath,braket}
\usepackage{color,calc,graphicx}
\usepackage[usenames,dvipsnames,svgnames,table,cmyk,hyperref]{xcolor}
\usepackage[colorlinks]{hyperref}
\hypersetup{
	colorlinks = true,
	citecolor = {blue},
	linkcolor = {blue},
	urlcolor = {blue}
}
\usepackage{graphicx}
\usepackage{amsmath}
\usepackage{latexsym}
\usepackage{bbm}

\newcommand{\degree}{{}^{\circ}}
\usepackage[charter,cal=cmcal,sfscaled=false]{mathdesign}
\usepackage{booktabs}
\usepackage{multirow}
\usepackage{subfigure}
\usepackage{dcolumn}
\usepackage{mathrsfs}

\usepackage{ulem}
\newcommand{\minew}[1]{{\color{black}{#1}}}
\newcommand{\miold}[1]{\iffalse{#1}\fi}
\begin{document}

\title{Loss-tolerant all-photonic quantum repeater with generalized Shor code}

\author{Rui Zhang}
\altaffiliation{These authors contributed equally to this work.}
\affiliation{Hefei National Laboratory for Physical Sciences at the Microscale and Department of Modern Physics, University of Science and Technology of China, Hefei 230026, China}
\affiliation{Shanghai Branch, CAS Center for Excellence in Quantum Information and Quantum Physics, University of Science and Technology of China, Shanghai 201315, China}
\affiliation{Shanghai Research Center for Quantum Sciences, Shanghai 201315, China}

\author{Li-Zheng Liu}
\altaffiliation{These authors contributed equally to this work.}
\affiliation{Hefei National Laboratory for Physical Sciences at the Microscale and Department of Modern Physics, University of Science and Technology of China, Hefei 230026, China}
\affiliation{Shanghai Branch, CAS Center for Excellence in Quantum Information and Quantum Physics, University of Science and Technology of China, Shanghai 201315, China}
\affiliation{Shanghai Research Center for Quantum Sciences, Shanghai 201315, China}

\author{Zheng-Da Li}
\altaffiliation{These authors contributed equally to this work.}
\affiliation{Hefei National Laboratory for Physical Sciences at the Microscale and Department of Modern Physics, University of Science and Technology of China, Hefei 230026, China}
\affiliation{Shanghai Branch, CAS Center for Excellence in Quantum Information and Quantum Physics, University of Science and Technology of China, Shanghai 201315, China}
\affiliation{Shanghai Research Center for Quantum Sciences, Shanghai 201315, China}

\author{Yue-Yang Fei}
\affiliation{Hefei National Laboratory for Physical Sciences at the Microscale and Department of Modern Physics, University of Science and Technology of China, Hefei 230026, China}
\affiliation{Shanghai Branch, CAS Center for Excellence in Quantum Information and Quantum Physics, University of Science and Technology of China, Shanghai 201315, China}
\affiliation{Shanghai Research Center for Quantum Sciences, Shanghai 201315, China}

\author{Xu-Fei Yin}
\affiliation{Hefei National Laboratory for Physical Sciences at the Microscale and Department of Modern Physics, University of Science and Technology of China, Hefei 230026, China}
\affiliation{Shanghai Branch, CAS Center for Excellence in Quantum Information and Quantum Physics, University of Science and Technology of China, Shanghai 201315, China}
\affiliation{Shanghai Research Center for Quantum Sciences, Shanghai 201315, China}

\author{Li Li}
\affiliation{Hefei National Laboratory for Physical Sciences at the Microscale and Department of Modern Physics, University of Science and Technology of China, Hefei 230026, China}
\affiliation{Shanghai Branch, CAS Center for Excellence in Quantum Information and Quantum Physics, University of Science and Technology of China, Shanghai 201315, China}
\affiliation{Shanghai Research Center for Quantum Sciences, Shanghai 201315, China}

\author{Nai-Le Liu}
\affiliation{Hefei National Laboratory for Physical Sciences at the Microscale and Department of Modern Physics, University of Science and Technology of China, Hefei 230026, China}
\affiliation{Shanghai Branch, CAS Center for Excellence in Quantum Information and Quantum Physics, University of Science and Technology of China, Shanghai 201315, China}
\affiliation{Shanghai Research Center for Quantum Sciences, Shanghai 201315, China}

\author{Yingqiu Mao}
\affiliation{Hefei National Laboratory for Physical Sciences at the Microscale and Department of Modern Physics, University of Science and Technology of China, Hefei 230026, China}
\affiliation{Shanghai Branch, CAS Center for Excellence in Quantum Information and Quantum Physics, University of Science and Technology of China, Shanghai 201315, China}
\affiliation{Shanghai Research Center for Quantum Sciences, Shanghai 201315, China}

\author{Yu-Ao Chen}
%\email{yuaochen@ustc.edu.cn}
\affiliation{Hefei National Laboratory for Physical Sciences at the Microscale and Department of Modern Physics, University of Science and Technology of China, Hefei 230026, China}
\affiliation{Shanghai Branch, CAS Center for Excellence in Quantum Information and Quantum Physics, University of Science and Technology of China, Shanghai 201315, China}
\affiliation{Shanghai Research Center for Quantum Sciences, Shanghai 201315, China}

\author{Jian-Wei Pan}
\affiliation{Hefei National Laboratory for Physical Sciences at the Microscale and Department of Modern Physics, University of Science and Technology of China, Hefei 230026, China}
\affiliation{Shanghai Branch, CAS Center for Excellence in Quantum Information and Quantum Physics, University of Science and Technology of China, Shanghai 201315, China}
\affiliation{Shanghai Research Center for Quantum Sciences, Shanghai 201315, China}

\begin{abstract}

The all-photonic quantum repeater (APQR) is a promising repeater scheme to realize long-distance quantum communication. 
For a practical APQR, an indispensable requirement is the robustness of the repeater graph state (RGS) against photon loss. 
We propose a new loss-tolerant scheme by applying the generalized Shor code to RGS, which can be experimentally demonstrated with current technology. 
Experimentally, we first prepare and verify the nine-qubit Shor code. 
Then, by applying the generalized Shor code to APQR and \minew{preparing a simplified encoded RGS with the structure of $1\times2$ based on the Shor code state}, the effectiveness of our loss-tolerant scheme and the loss-tolerance of the encoded RGS are respectively verified.
Our results make an essential step towards the practical APQR and enrich the research of quantum error correction code.
\end{abstract}

\maketitle

Quantum repeaters, which connect short-distance communication nodes segment by segment, are  essential components  in wide-area quantum networks \cite{PhysRevLett.81.5932,duan2001long}. 
The all-photonic quantum repeater (APQR), which can eliminate the necessity for quantum memories by connecting repeater nodes with all-photonic repeater graph states (RGSs), has been proven theoretically \cite{azuma2015all,ratePhysRevA.95.012304,buildingPhysRevA.100.052303} and experimentally \cite{hasegawa2019experimental,li2019experimental} a promising scheme to achieve long-distance quantum communication. 
Due to its obvious advantages, it is with great significance to verify the feasibility of APQR in realistic applications, particularly noisy environments, which require RGS to be robust against photon loss \cite{azuma2015all,ratePhysRevA.95.012304,buildingPhysRevA.100.052303}. 
In previous proof-of-principle demonstrations, the Greenberger-Horne-Zeilinger (GHZ) state is used as the RGS (bare-GHZ RGS)\cite{hasegawa2019experimental,li2019experimental}, which may break down the repeater scheme since the GHZ state is not loss-tolerant. 
In the original APQR scheme \cite{azuma2015all}, a great number of tree-like graph states are used to protect RGS, while it is resource-consuming and challenging to realize in the short term \cite{PhysRevLett.97.120501,ratePhysRevA.95.012304,buildingPhysRevA.100.052303}.
Thus a practical loss-tolerant scheme to protect APQR from photon loss is missing up to now.

To protect quantum information from environmental noise, quantum error correction codes (QECCs) play an essential role in many areas of fault-tolerant quantum information processing \cite{Phys.Rev.A1995ShorScheme,PhysRevLett.81.2152,chiaverini2004realization,stricker2020experimental,waldherr2014quantum, reed2012realization,PhysRevA.71.052332,yao2012experimental}. 
The quantum parity code (QPC) \cite{Loss-TolePhysRevLett.95.100501}, due to its powerful robustness against physical qubit loss, has been widely studied to protect quantum states \cite{PhysRevA.77.012310,Lu11050,PhysRevLett.112.080801,luo2020quantum}, especially in quantum networks \cite{munro2012quantum,PhysRevLett.112.250501,buildingPhysRevA.100.052303,PhysRevLett.117.210501,staticPhysRevA.95.012327}.
As the representative of QPC, the nine-qubit Shor code has a clear design of a combination of the three-qubit bit-flip and phase-flip codes, and encodes a single-qubit state $\ket{\Phi}=\alpha\ket{0}+\beta\ket{1}$ into the logical state 
\begin{equation}
\begin{aligned}	
\ket{\Phi}_{l}=\alpha(\ket{000}+\ket{111})^{\otimes3}+\beta(\ket{000}-\ket{111})^{\otimes3},\\	
\end{aligned}
\label{Eq:Shorcodestate}
\end{equation}
where $(\ket{000}\pm\ket{111})$ are code blocks \cite{Phys.Rev.A1995ShorScheme,2010NielsenQuantum}. If a physical qubit is lost, the loss-affected code block will become a mixed one, but the influence is local and limited to that code block only. 
As long as there are photons remained in the loss-affected code block, the overall entanglement structure of the three code blocks will not be destroyed, and the entanglement among physical qubits in the remaining two intact code blocks will survive \cite{Lu11050,munro2012quantum}.
Thus, benefitting from strong loss-tolerance, the nine-qubit Shor code can be used to protect the bare-GHZ RGS \cite{hasegawa2019experimental,li2019experimental} from photon loss.
Specifically, since the three code blocks of the nine-qubit Shor code is arranged in the GHZ-type structure, it is natural to apply them to the bare-GHZ RGS and encode every qubit of the GHZ state into a logical qubit with a code block, i.e., $\ket{0} \mapsto \ket{0}_{l}=\frac{1}{\sqrt{2}}(\ket{000}+\ket{111})$ and $\ket{1} \mapsto \ket{1}_{l}=\frac{1}{\sqrt{2}}(\ket{000}-\ket{111})$.
Thus, the bare-GHZ RGS is encoded into the nine-qubit Shor code state and equipped with strong loss-tolerance. 
In this way, the encoded RGS with $n$ logical qubits can be viewed as a generalized Shor code state with $n$ code blocks.
The influence of photon loss would then be limited to the given loss-affected logical qubit, which will not destroy the overall entanglement links of the encoded RGS.
Also, since it is encoded by a full-featured QECC, the encoded RGS can be naturally endowed with robustness against arbitrary single-qubit errors \cite{Phys.Rev.A1995ShorScheme,2010NielsenQuantum}.

In this work, by manipulating a ten-photon interferometer, we prepare the nine-qubit Shor code and investigate the feasibility of our loss-tolerant scheme for APQR. 
To verify the prepared nine-qubit Shor code, we encode and read out six different states, and verify its robustness to qubit loss and capability of identifying arbitrary single-qubit errors.
Then, we apply QPC to the bare-GHZ RGS to prepare a partially encoded RGS and perform a demonstration of the entanglement connection process for the most simplified loss-tolerant APQR. 
Thus, the effectiveness of our loss-tolerant proposal is verified.
Furthermore, based on the prepared nine-qubit Shor code, we investigate the loss-tolerance of the encoded RGS by measuring the entanglement between two intact logical qubits when photon loss occurs in the third logical qubit. 
Our results indicate the entanglement between the two logical qubits can be well maintained in a loss-affected encoded RGS, and thus the loss-tolerance of the encoded RGS is also verified.

\begin{figure*}
	\centering
	\includegraphics[width=1 \linewidth]{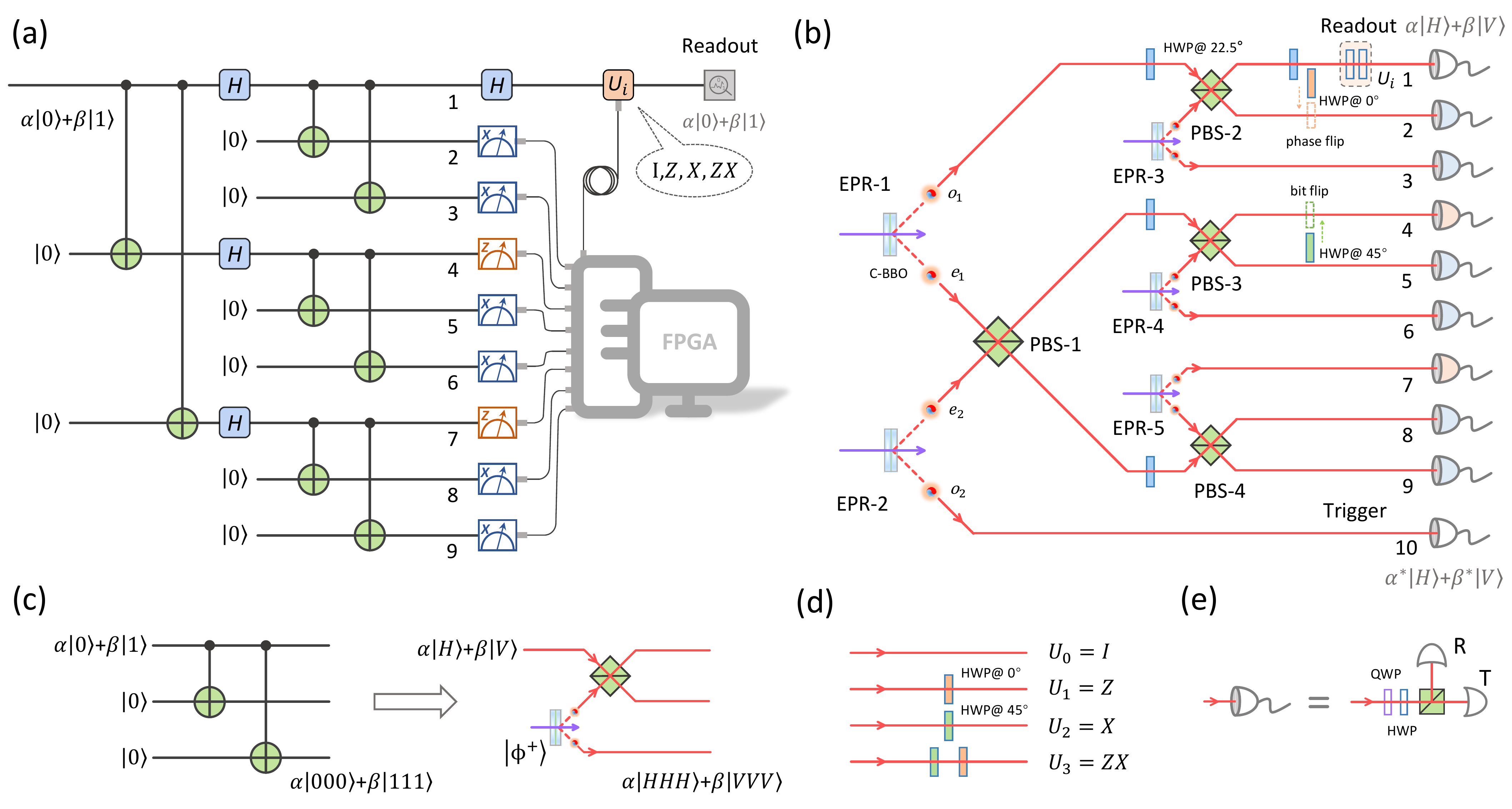}
	\caption{\textbf{Schematic diagrams for the circuit of the nine-qubit Shor code and its linear optical realization.}
		(a) The quantum circuit for encoding and readout of the nine-qubit Shor code. 
		The measurement in yellow indicates measurement in $Z$ basis, $M_{Z}$. 
		The measurement in blue indicates measurement in $X$ basis, $M_{X}$.
		(b) The experimental setup for the nine-qubit Shor code. 
		Flip errors are realized with half-wave plates (HWPs).
		The X error is introduced to photon 4 with an HWP at $45\degree$.
		The Z error is introduced to the first code block with an HWP at $0\degree$ in the path of photon 2.
		(c) Quantum encoder for preparing the code block and its experimental realization.
		(d) Experimental settings of unitary operations $U_{i}$.
		(e) Detailed schematics for the measurement device. 
		The measurement device in (b) represents a polarization analyzer, which is composed of a quarter-wave plate (QWP), an HWP, a polarizing beam splitter (PBS), and two single-photon detectors. 	
		All detector output signals are sent into a field-programmable gate arrays (FPGA) coincidence unit, and tenfold coincidence events are post-selected.}
	\label{Fig:recoverycircuit}
\end{figure*}

\begin{figure*}[htbp!]
	\centering
	\includegraphics[width=0.9 \linewidth]{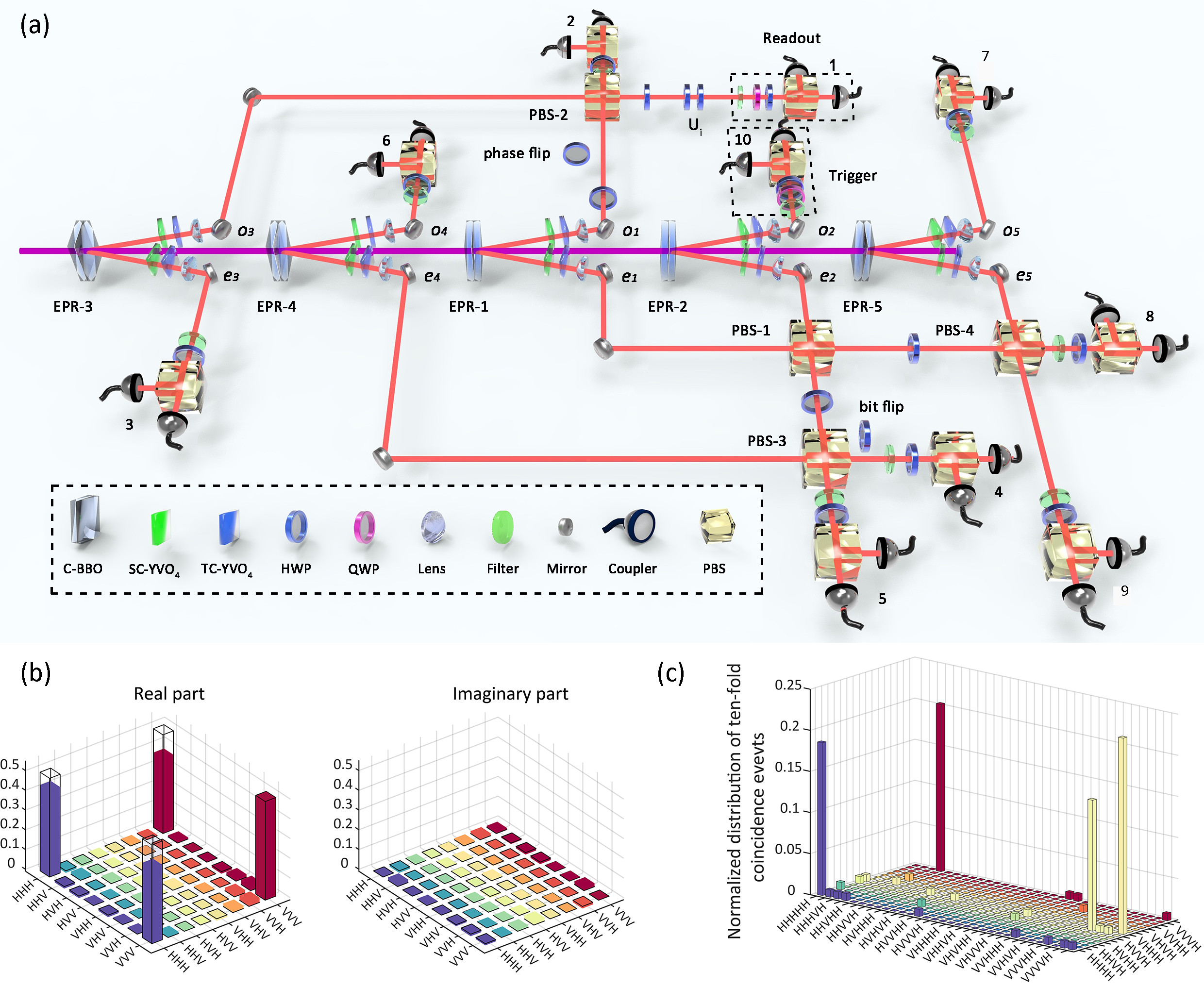}
	\caption{\textbf{Experimental setup and its characterizations.}
		(a) Schematic diagram of the experimental setup. The order of entanglement sources is arranged according to the circuit in Fig. 1b. The $\text{SC-YVO}_{4}$ and $\text{TC-YVO}_{4}$ respectively represent spatial and temporal compensation yttrium orthovanadate crystal.
		(b) Reconstructed density matrix of the code block $\rho^{ex}$ in the first quantum encoder. The empty and boxes respectively denote the ideal and experimental results.
		(c) Polarization distribution of the logical Shor code state $\ket{D}_{l}$ in $\ket{H/V}$ basis.
	}
	\label{Fig:setup}
\end{figure*}

\section{Experimental setup}

\begin{figure*}[htbp!]
	\centering
	\includegraphics[width=1 \linewidth]{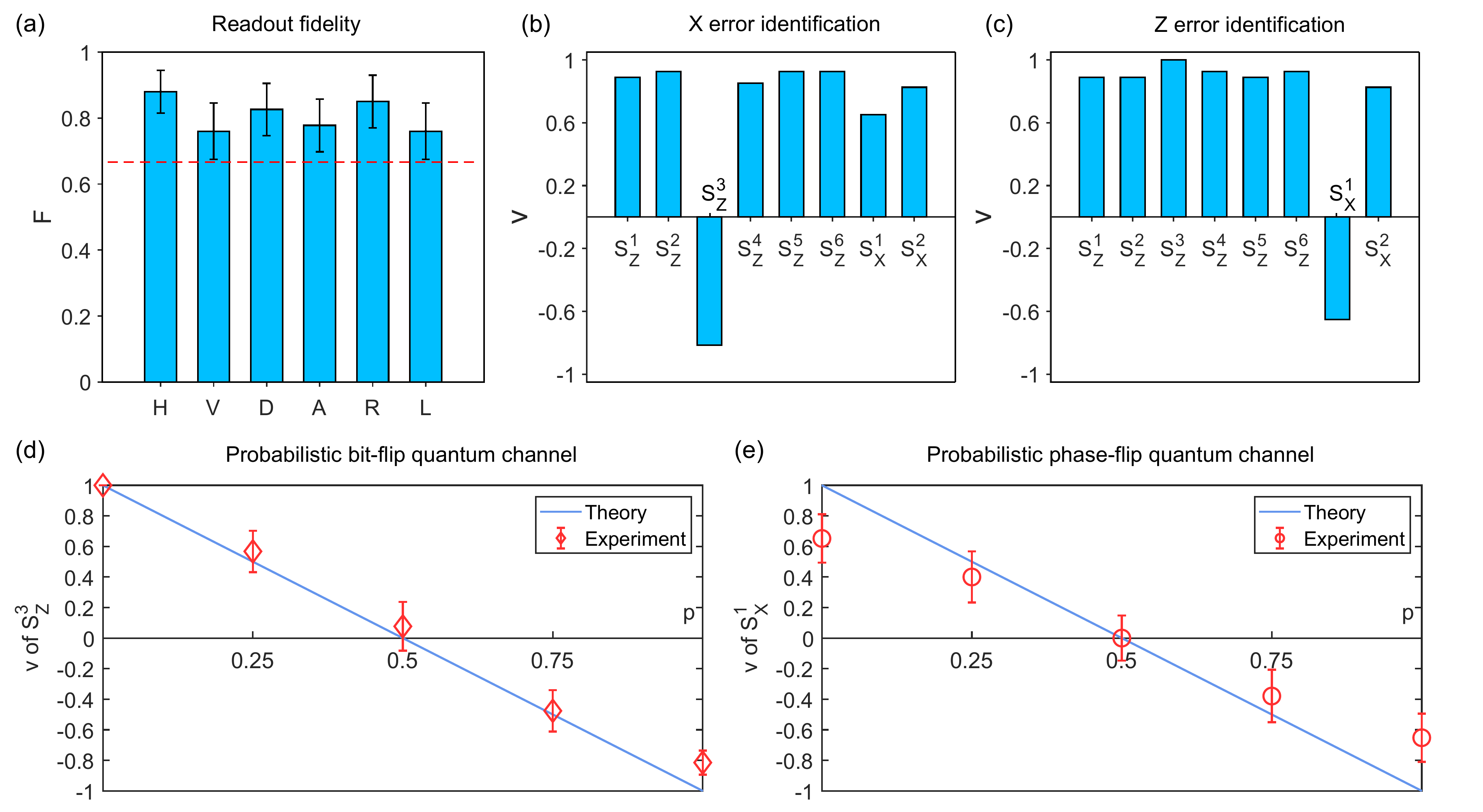}
	\caption{\textbf{Performance of the prepared nine-qubit Shor code.}
		(a) Readout fidelities for six different encoded states. 
		(b) Visibilities of syndrome diagnosis operators for the X error on the physical qubit $4$.
		(c) Visibilities of syndrome diagnosis operators for the Z error in the first code block.
		(d) Visibilities of the probe syndrome diagnosis operators $S_{Z}^{3}$ with respect to error probabilities $p$ for the bit-flip quantum channel.
		(e) Visibilities of the probe syndrome diagnosis operators $S_{X}^{1}$ with respect to error probabilities $p$ for the phase-flip quantum channel.		
	}
	\label{Fig:error}
\end{figure*}

\begin{figure*}[htbp!]
	\centering
	\includegraphics[width=1 \linewidth]{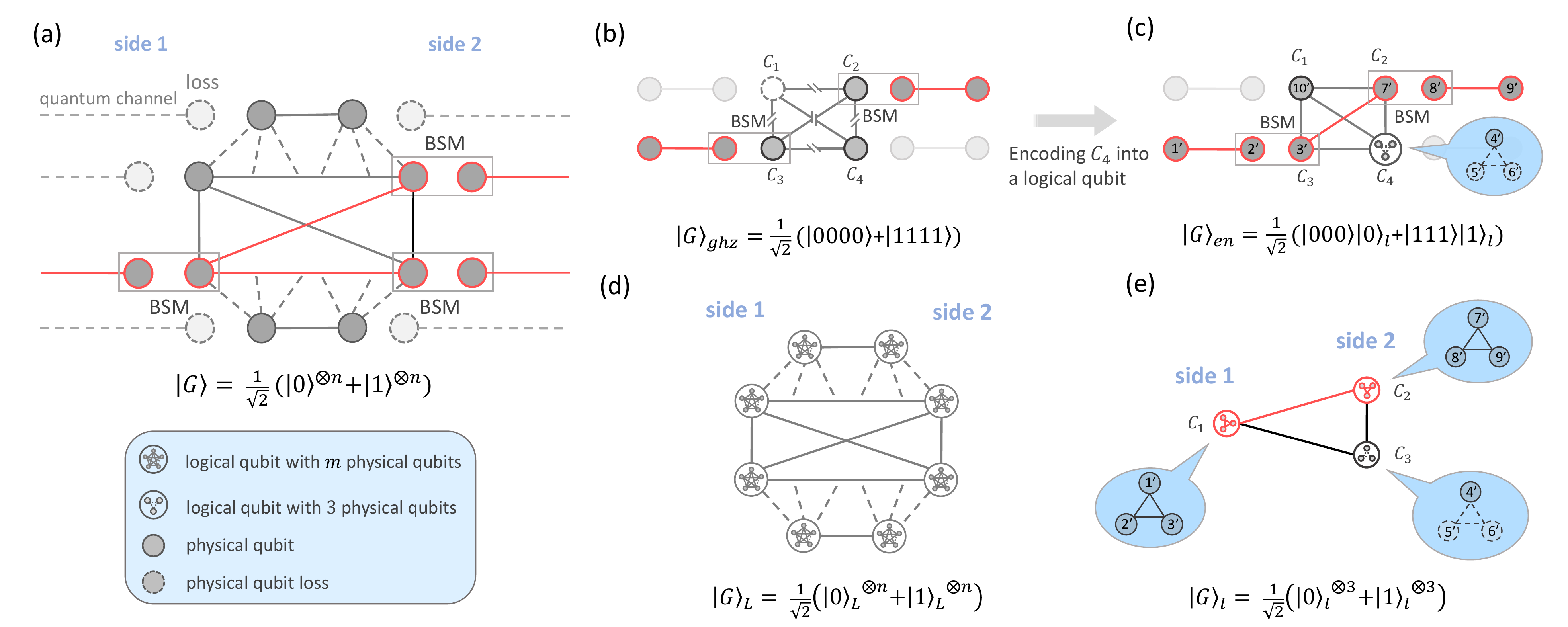}
	\caption{\textbf{Schematic diagrams for APQR and encoded RGS.}
		(a) The APQR scheme with the bare-GHZ RGS $\ket{G}$.
		Any BSM that succeeds on each side of the RGS heralds entanglement connection between two corresponding channels (indicated by the red line), and thus the APQR provides an exponential improvement in connection probability compared with the parallel entanglement swapping.
		(b) The sketch of APQR based on the bare-GHZ RGS. Once the photon loss occurs, the bare-GHZ RGS would become a mixed state and the APQR fails.	
		(c) The most simplified APQR with a partially encoded RGS $\ket{G}_{en}=\frac{1}{\sqrt{2}}(\ket{000}\ket{0}_{l}+\ket{111}\ket{1}_{l})$. 
		The entanglement connection of two quantum channels on two sides of the partially encoded RGS can still be established when photon(s) loss occurs in the logical qubit $C_{4}$. 
		(d) The loss-tolerant RGS $\ket{G}_{L}$ encoded with the generalized Shor code. 
		\minew{(e) The simplified encoded RGS $\ket{G}_{l}$ with $m=3$. The loss of one or two photons in the logical qubit $C_{3}$ does not affect the entanglement between two intact logical qubits $C_{1}$ and $C_{2}$, reflecting the strong loss-tolerance of the encoded RGS.}					
	}
	\label{Fig:RGS}
\end{figure*}

We verify the feasibility of the loss-tolerant scheme for APQR based on the nine-qubit Shor code.
Figure \ref{Fig:recoverycircuit}(a) shows the quantum circuit for preparing the nine-qubit Shor code of Eq.~(\ref{Eq:Shorcodestate}).
First, the input state to be encoded is designated as the common control state of two controlled-NOT (CNOT) gates with the fixed target qubits $\ket{0}$, and the code block $(\alpha\ket{000}+\beta\ket{111})$ can be prepared with this quantum circuit, which is called a ``quantum encoder''.
Then, each of these three qubits is respectively rotated by a Hadamard ($H_d$) gate and subsequently designated as the control qubit of three other quantum encoders to encode three more code blocks.
In this way, after four quantum encoders, the nine-qubit Shor code is prepared.
In our experiment, by encoding logical qubits $\ket{0}$ and $\ket{1}$ with the horizontal ($\ket{H}$) and vertical ($\ket{V}$) polarizations of photons, the CNOT gate for the specific target qubit $\ket{H}$ can be simulated by a polarizing beam splitter (PBS) and a half-wave plates (HWP) at $22.5\degree$ with a success probability of 50$\%$ after post-selection \cite{PanRMP2012,Lu11050}.
As shown in Fig.~\ref{Fig:recoverycircuit}(c), the quantum encoder with two such CNOT gates can be further simulated by interfering the control state $(\alpha\ket{H}+\beta\ket{V})$ with an ancillary Einstein-Podolsky-Rosen (EPR) pair $\ket{\Phi^+}=\frac{1}{\sqrt{2}}(\ket{HH}+\ket{VV})$ on a PBS.
Thus, via the linear optical circuit as shown in Fig.~\ref{Fig:recoverycircuit}(b), the nine-qubit Shor code is prepared.
By triggering photon $o_2$ from the entanglement source EPR-2 to the state $\ket{\phi}_{t}=\alpha^{*}\ket{H}_{10}+\beta^{*}\ket{V}_{10}$, the twin photon $e_2$ with the initial state $\ket{\phi}_{i}=\alpha\ket{H}+\beta\ket{V}$ is injected into the first quantum encoder, and the corresponding nine-qubit Shor code is prepared with three more quantum encoders.
For a special logical state of the nine-qubit Shor code
\begin{equation}
\ket{D}_{l}=\frac{1}{4}[(\ket{HHH}+\ket{VVV})^{\otimes3}+(\ket{HHH}-\ket{VVV})^{\otimes3}],
\label{Eq:DShorcodestate}
\end{equation}
demonstrations of identifying single-qubit errors and verifying the loss-tolerance of the encoded RGS are to be implemented based on it.
The readout of encoded states from the nine-qubit Shor code is implemented by analyzing results of collective measurement on photons $2\sim9$ and applying suitable unitary operations of $U_{i}\in\{I, Z, X, ZX\}$ to the photon 1 ($I, X, Y, Z$ are Pauli operators). 
The settings of $U_{i}$ and the measurement device are shown in Figs.~\ref{Fig:recoverycircuit}(d) and \ref{Fig:recoverycircuit}(e), respectively.

The actual experimental setup is depicted in Fig.~\ref{Fig:setup}(a). 
A beam of ultraviolet pulsed laser (390 nm, 150 fs, and 80 MHz), is consecutively focused on five separate sandwich-like combinations of the $\beta$-barium borate (C-BBO) crystal and generate five EPR pairs $\ket{\Phi^+}$ via type-II spontaneous parametric down-conversion (SPDC) processing.
Then photons from different EPR sources overlap on PBSs to prepare code blocks.
With pump power of $800~\text{mW}$ and filters of $\Delta\lambda= 3.6$~nm and $\Delta\lambda= 8$~nm respectively on extraordinary ($e$) and ordinary ($o$) photons, the overall collection efficiency of a pair of photons from an EPR source $\eta$ is $38\%$, the average down-conversion probability for each of the five EPR sources $P$ is 0.06, and the average interference visibility of PBSs in four quantum encoders $V$ is $0.70$.
We implement the performance test of the experimental setup. 
First, by projecting the trigger photon $o_{2}$ into the state $\ket{\phi}_{t}=\frac{1}{\sqrt{2}}(\ket{H}_{10}+\ket{V}_{10})$ and interfering photons $e_{1}$ and $e_{2}$ on the PBS-1 of the first quantum encoder, we perform state tomography on the first code block state $\ket{\Phi}^{id}=\frac{1}{\sqrt{2}}(\ket{HHH}+\ket{VVV})$. 
%The photons $e_{1}$, $o_{1}$, and $e_{2}$ are then measured in 27 measurement settings and followed by data processing with the maximum likelihood method \cite{PhysRevA.64.052312}. 
The reconstructed density matrix $\rho^{ex}$ is shown in Fig.~\ref{Fig:setup}(b), and the fidelity of the code block can be calculated as $F=~^{id}\bra{\Phi}\rho^{ex}\ket{\Phi}^{id}=0.92$.
Afterward, the nine-qubit Shor code state $\ket{D}_{l}$ in Eq. (\ref{Eq:DShorcodestate}) is prepared with three more quantum encoders, and we measure it in the $\ket{H/V}$ basis to verify its polarization distribution. 
The results are illustrated in Fig.~\ref{Fig:setup}(c), where the four components in the state vector are dominant and the total signal-to-noise ratio (SNR) is $3.71:1$.
Overall, the quality of our experimental setup is adequately high for subsequent demonstrations.

\section{Results}

\subsection{Performance of the prepared nine-qubit Shor code}

As the core tool for verifying our scheme, we first carry out a verification demonstration of the prepared nine-qubit Shor code, including reading out encoded states, tolerating physical qubit(s) loss, and identifying arbitrary single-qubit errors.

First, we choose six different input states $\ket{\phi}_{i}$ from the complete orthogonal basis set $\{\ket{H},\ket{V},\ket{D},\ket{A},\ket{R},\ket{L}\}$ with $\ket{D/A}=\frac{1}{\sqrt{2}}(\ket{H}\pm\ket{V})$ and $\ket{R/L}=\frac{1}{\sqrt{2}}(\ket{H}\pm i\ket{V})$, and encode each of them into the nine-qubit Shor codes.
The corresponding readout fidelities are shown in Fig.~\ref{Fig:error} (a).
Each readout fidelity for the six encoded states exceeds the classical limit $2/3$, and the average fidelity is as high as $0.81 \pm 0.03$, which fully reflects that the nine-qubit Shor code we prepared works for a general unknown state.

Next, by removing polarization analysis components (QWP, HWP, and PBS) from the measurement device and letting the photon directly enter the detector without knowing its polarization information, the loss of physical qubit in the Shor code can be simulated. 
On this basis, we demonstrate that the encoded states can still be read out even if one or two physical qubits are lost \cite{Lu11050}.
Experimentally, two encoded states $\ket{D}$ and $\ket{A}$ are retrieved from the nine-qubit Shor codes with high fidelities of $0.80 \pm 0.08$ and $0.76 \pm 0.08$  respectively when the physical qubit 6 is lost, and $0.88 \pm 0.06$ and $0.76 \pm 0.08$  respectively when the physical qubits 4 and 6 are both lost.
\minew{Since the quantum information is encoded by redundant coding, reducing imperfect redundant physical qubit(s) also reduces noise in the state space of the logical qubit, and thus the fidelity of the logical qubit state which contains quantum information would increase.}
The readout fidelities of the encoded states are not affected by the physical qubit(s) loss, reflecting the strong loss-tolerance of the nine-qubit Shor code.

Then, based on the fact that an arbitrary single-qubit error in the nine-qubit Shor code can collapse into the discrete single-qubit bit-flip and/or phase-flip errors (Corresponding to the X and Z errors) with the eight syndrome diagnosis operators, i.e., the eight stabilizers, $S_{Z}^{1}=Z_{1}Z_{2},~S_{Z}^{2}=Z_{2}Z_{3}; 
~S_{Z}^{3}=Z_{4}Z_{5}, 
~S_{Z}^{4}=Z_{5}Z_{6};
~S_{Z}^{5}=Z_{7}Z_{8},
~S_{Z}^{6}=Z_{8}Z_{9}$ and 
$S_{X}^{1}=X_{1}X_{2}X_{3}X_{4}X_{5}X_{6},
~S_{X}^{2}=X_{4}X_{5}X_{6}X_{7}X_{8}X_{9}$.
We focus on identifying the single-qubit X and Z errors to investigate its performance of identifying single-qubit errors \cite{2010NielsenQuantum,aoki2009quantum}.
Experimentally, these two flip errors are artificially introduced to physical qubits by inserting HWPs into light paths as shown in Fig.~\ref{Fig:recoverycircuit}(b), and identified by applying measurements on syndrome diagnosis operators.
The visibilities of each operator for two cases are respectively illustrated in bar charts as shown in Figs.~\ref{Fig:error}(b) and \ref{Fig:error}(c).
The negative values of $S_{Z}^{3}$ and $S_{X}^{1}$ in the two charts fully verify that flip errors respectively happen to photon $4$ with the X error and the first code block with the Z error, exactly as we introduced.
Furthermore, we simulate a more general and practical case that physical qubits pass through a probabilistic bit-flip (phase-flip) quantum channel, in which they may suffer from the X (Z) error with various probabilities $p$ = 0, 0.25, 0.5, 0.75, and 1.
Visibilities for the probe syndrome diagnosis operators, $S_{Z}^{3}$ for the bit-flip channel and $S_{X}^{1}$ for the phase-flip channel, with respect to various error probabilities $p$ are illustrated in Figs. \ref{Fig:error}(d) and \ref{Fig:error}(e). 
With experimental results overall in close accordance with theory, the capability of the nine-qubit Shor code to identify single-qubit errors is demonstrated (the full list of visibilities for all syndrome diagnosis operators can be found in Supplemental document).

\begin{figure}[htbp!]
	\centering
	\includegraphics[width=1 \linewidth]{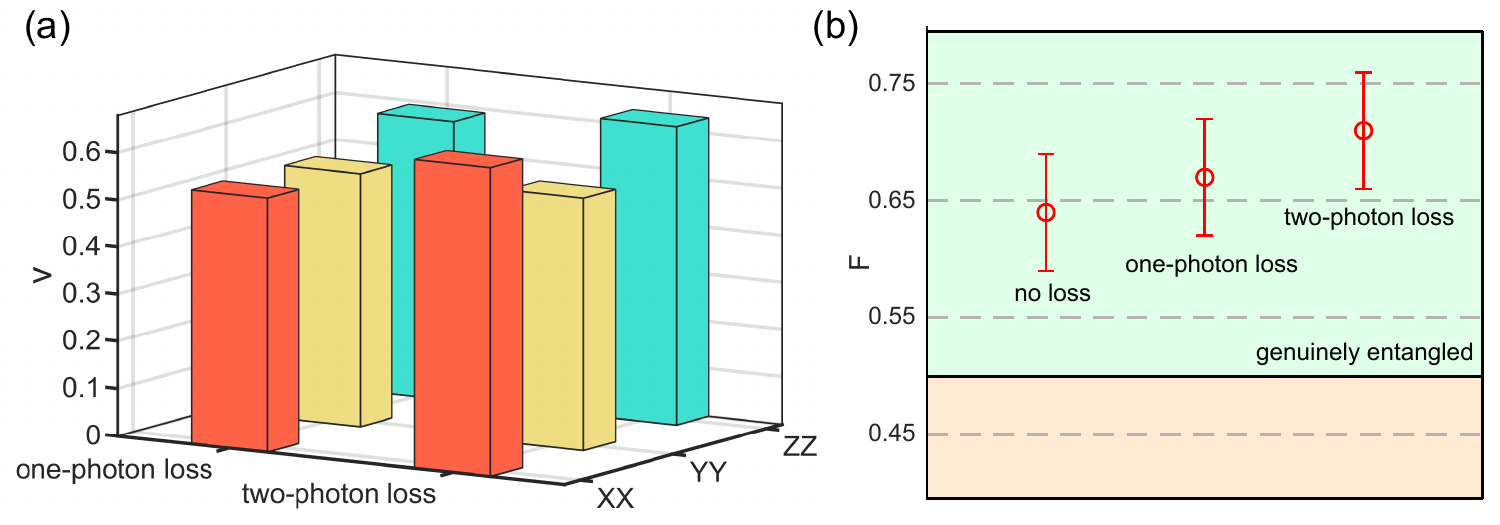}
	\caption{ 
		\textbf{Results for demonstration of the loss-tolerant APQR scheme.}	
		(a) Results for verification of the effectiveness of the loss-tolerant proposal.
		(b) Results for investigation of the loss-tolerance for the encoded RGS $\ket{G}_{l}$.
	}
	\label{Fig:RGSdata}
\end{figure}

\subsection{Demonstration of the loss-tolerant APQR scheme}

In the APQR scheme, as shown in Fig.~\ref{Fig:RGS}(a), a specially designed graph state---the RGS, is used as a flexible switch to connect multi-parallel quantum channels between two adjacent nodes in the quantum network \cite{azuma2015all}.
For each of two sides of the RGS, any one of $n$ connection operations, such as the Bell state measurements (BSMs), on photons from channels and those in the RGS succeeds, heralding the entanglement connection between the two corresponding quantum channels.
In such a way, without requiring quantum memories, the APQR provides an exponential improvement in connection probability compared with conventional multi-parallel and single-connected channels \cite{azuma2015all,hasegawa2019experimental,li2019experimental}.
In practice, the all-photonic RGS will inevitably suffer from photon loss due to environmental noises, which would destroy the unprotected bare-GHZ RGS and fail the APQR scheme as shown in Fig.~\ref{Fig:RGS}(b).
We propose, based on the bare-GHZ RGS  ($\ket{G}=\frac{1}{\sqrt{2}}(\ket{0}^{\otimes n}+\ket{1}^{\otimes n})$) in the previous researches \cite{hasegawa2019experimental,li2019experimental}, to apply QPC to it and encode every qubit of it into a logical qubit with $m$ physical qubits. 
That is,
\begin{equation}
\begin{aligned}
\ket{G} \mapsto \ket{G}_{L}=\frac{1}{\sqrt{2}}(\ket{0}_{L}^{\otimes n}+\ket{1}_{L}^{\otimes n}),\\
\end{aligned}
\label{Eq:encoding}
\end{equation}
with $\ket{0}_{L}=\frac{1}{\sqrt{2}}(\ket{0}^{\otimes m}+\ket{1}^{\otimes m})$ and $\ket{1}_{L}=\frac{1}{\sqrt{2}}(\ket{0}^{\otimes m}-\ket{1}^{\otimes m})$ as shown in Fig.~\ref{Fig:RGS}(d).
The encoded RGS can thus be considered as a generalized Shor code state and has a similar loss-tolerance as the nine-qubit Shor code.
In this work, we experimentally perform proof-of-principle demonstrations of this loss-tolerant APQR scheme.

By encoding a qubit of the bare-GHZ RGS $\ket{G}_{ghz}=\frac{1}{\sqrt{2}}(\ket{0000}+\ket{1111})$ into the logical qubit as shown in Fig.~\ref{Fig:RGS}(c), we demonstrate the entanglement connection process of a simplified APQR with the partially encoded RGS  $\ket{G}_{en}=\frac{1}{\sqrt{2}}(\ket{000}\ket{0}_{l}+\ket{111}\ket{1}_{l})$ with $m=3$ in our experimental setup.
\minew{Considering the case of photon(s) loss occurring in the logical qubit $C_{4}$, we investigate the entanglement connection of the repeater network by executing the entanglement witness (EW) \cite{guhne2009entanglement} on the terminal photons $1'$ and $9'$. 
	Experimentally, by measuring photon $10'$ of logical qubit $C_{1}$ in $X$ basis, and the surviving photon(s) of logical qubit $C_{4}$ in $Z$ basis, followed by performing BSMs on photons $2'$ and $8'$ from quantum channels with photons $3'$ and $7'$ of the partially encoded RGS, the entanglement between photons $1'$ and $9'$ can be witnessed by measuring them in $XX$, $YY$, and $ZZ$ bases.
	The experimentally measured fractions are shown in Fig.~\ref{Fig:RGSdata}(a), and the expected values of EW are quantified to be $W_{1} = -0.17\pm 0.05$ for the one-photon loss case, and $W_{2}=-0.21 \pm0.05$ for the two-photon loss case. 
	Both of which clearly indicate the final shared state between the two terminal photons is entangled \cite{guhne2009entanglement}.
	In this way, the effectiveness of our loss-tolerant proposal for APQR is verified.}

The loss-tolerance of the APQR comes from the encoded RGS, and the requirements for an encoded RGS are (i) at least one intact logical qubit on each side of the encoded RGS, and (ii) at least one photon remained in each loss-affected logical qubit of the encoded RGS \cite{munro2012quantum,buildingPhysRevA.100.052303,ratePhysRevA.95.012304}.
As the core component of the loss-tolerant APQR, the examination of the loss-tolerant performance for the encoded RGS is necessary. 
\minew{The most simplified encoded RGS $\ket{G}_l$ with $n=3$ and $m=3$ in Fig.~\ref{Fig:RGS}(e) can be prepared in our experimental setup, which is exactly the nine-qubit Shor code state $\ket{D}_l$ in Eq. (\ref{Eq:DShorcodestate}).
	Its loss-tolerance can be investigated by observing the variation of entanglement fidelity between two photons in two intact logical qubits $C_{1}$ and $C_{2}$ when photon(s) loss occurs in the third logical qubit $C_{3}$ \cite{guhne2009entanglement}.
	The density matrix of an ideal entanglement states shared between photons $1'$ and $9'$ can be decomposed as 
	$\ket{\Phi^+}\bra{\Phi^+}=\frac{1}{4}(I+XX-YY+ZZ)$.
	Thus, by measuring the remaining photons $2'$, $3'$ and $7'$, $8'$ of $C_{1}$ and $C_{2}$ in $X$ basis and the surviving photon(s) of $C_{4}$ in $Z$ basis, the fidelity of shared state between photons $1'$ and $9'$ can be determined by measuring in $XX$, $YY$ and $ZZ$ bases.
	The overall fidelities are respectively quantified to be  $0.64\pm 0.05$ for the no loss case, $0.67\pm 0.05$ for the one-photon loss case, and $0.71\pm 0.05$ for the two-photon loss case, as shown in Fig.~\ref{Fig:RGSdata}(b). 
	The entanglement fidelity of two photons of two intact logical qubits is well maintained when photon(s) loss continuously occurs in the third logical qubit, which is a good reflection of the effectiveness of entanglement protection of our loss-tolerant scheme for encoded RGS.}

\section{Discussion}

In this work, we have proposed a new loss-tolerant scheme for the APQR and studied its feasibility by implementing experimental demonstrations on the nine-qubit Shor code with a ten-photon interferometer. 
Firstly, we have verified the prepared nine-qubit Shor code, including the high-fidelity readout of encoded states, the robustness test for the physical qubit loss, and the accurate identification of single-qubit errors. 
\minew{Secondly, we have verified the effectiveness of the loss-tolerant scheme by demonstrating entanglement connection of the simplified loss-tolerant APQR based on the partially encoded RGS.
	Thirdly, we have prepared the most simplified encoded RGS with the structure of 1$\times$2 based on the Shor code state $\ket{D}_{l}$, through which we have demonstrated the loss-tolerance of the encoded RGS.}

It is a resource-efficient way to encode the RGS with the generalized Shor code. 
By encoding the RGS with such a symmetrical code, the number of photons consumed for encoding a logical qubit of an encoded RGS is the number of physical qubits of a code block (three in this work), while it will be the total number of physical qubits for the quantum code which is not in a GHZ-type structure, e.g., five for the $[[5,1,3]]$ code \cite{PhysRevLett.77.198}.
In addition, comparing with the original scheme \cite{azuma2015all}, our loss-tolerant scheme can achieve a comparable transmission rate with one order of magnitude fewer photons, when the concatenated Bell measurement (CBM) is employed to connect adjacent encoded RGSs (refer to Supplemental document or Ref. \cite{buildingPhysRevA.100.052303}).
\minew{In Ref. \cite{buildingPhysRevA.100.052303}, Lee \textit{et al}. have proposed the CBM as an efficient and resource-saving fundamental building block for all-optical scalable quantum networks. The CBM can also be performed to link the adjacent encoded RGS in our APQR network. In both quantum network protocols of Ref. \cite{buildingPhysRevA.100.052303} and ours, quantum information is encoded in the QPC. The quantum information is transmitted along the network in a one-way style in the former design, while the entanglement is distributed between remote places in our symmetric APQR network. Since CBM is the core means applied to QPC to connect repeater nodes in both designs of quantum networks, the overall communication efficiency and photon consumption are the same \cite{buildingPhysRevA.100.052303}.}
In this work, we have set the photon number per logical qubit of the encoded RGS to three, while it can be expanded to $m$ in realistic applications.
For a given transmission efficiency $\eta$, there exists the optimal ${n, m}$ for our loss-tolerant APQR to obtain the maximum communication efficiency and the minimum photon consumption.

Note that the probabilistic property of the SPDC source and post-selection of coincidence events cause the generation efficiency of multi-photon entangled states to decrease exponentially with the increase of photon number, which seriously limits the communication efficiency of our loss-tolerant APQR.
Recently, on-demand photon sources have been developed both in theory and experiments, which can be used to generate huge photonic graph states equipped with fusing operations \cite{Schwartz434,PhysRevX.7.041023,fusingPhysRevLett.95.010501,ratePhysRevA.95.012304,QuantumHilaire2021resource,DeterministicPhysRevLett.125.223601}. 
Without the need for post-selection on coincidence events, it is a very promising platform for generating large loss-tolerant encoded RGSs and implementing a practical APQR in the future.

\section{Funding} 
This work was supported by the National Key Research and Development (R$\&$D) Plan of China (2018YFA0306501), the National Natural Science Foundation of China (Grants No. 11975222, 11425417, 12104444, and U1738140), the Shanghai Municipal Science and Technology Major Project (2019SHZDZX01), the Anhui Initiative in Quantum Information Technologies, the Chinese Academy of Sciences and the Shanghai Science and Technology Development Funds (18JC1414700), and the Key (R$\&$D) Program of Guangdong province (Grant No. 2018B030325001). Y. M. acknowledges support from the China Postdoctoral Science Foundation (No. 2021M693093).
%%%%%%%%%%%%%%%%%%%%%%%%%%%%%%%%%%%%%%%%%%%%%%%%%%%%%%%%%%%%%%%%%%%

\bibliography{APQR}

%%%%%%%%%%%%%%%%%%%%%%%%%%%%%%%%%%%

\end{document}